\documentstyle[12pt]{article}

\begin{document}

\title{\Large\bf SUPERFIELD COVARIANT QUANTIZATION WITH BRST SYMMETRY}
\author{\sc P.M. Lavrov, P.Yu. Moshin\\
       {\small\it Tomsk State Pedagogical University, Tomsk 634041, Russia}}
\date{}

\maketitle

\begin{quotation}
\noindent
\footnotesize
 We generalize the method of superfield Lagrangian BRST quantization in the
 part of the gauge-fixing procedure and obtain a quantization method that
 can be considered as an alternative to the Batalin--Vilkovisky formalism.
\end{quotation}

\section{Introduction}
 In previous years it has been realized that two principles are essential
 in the formulation of appropriate quantum versions of a classical theory
 considered in the covariant, i.e.~Lagrangian, approach, namely, the
 invariance under the BRST transformations \cite{BRST}, and the
 gauge-independence of the $S$-matrix.

 It is well-known that the first principle can be replaced by the requirement
 of BRST--antiBRST symmetry \cite{BRSTantiBRST}, which also leads to possible
 versions of quantization methods, such as the $Sp(2)$-covariant quantization
 \cite{sp2}, the triplectic quantization \cite{tripl}, or the extended
 superfield BRST quantization \cite{extsupBRST}.

 For the first time, the two above-mentioned principles were realized in the
 framework of the BV quantization method \cite{BV} for general gauge theories,
 proposed by Batalin and Vilkovisky.

 It should be noted that the requirements of (extended) BRST symmetry and
 gauge-independence are two independent principles, which can be exemplified
 by the method of $Osp(1,2)$-covariant quantization \cite{osp} that satisfies
 the requirement of extended BRST symmetry, while violating the requirement
 of gauge-in\-de\-pen\-dence.

 Note also that there is arbitrariness in the realization of the above
 principles, which implies that there can be off-shell different quantum
 versions of a classical theory. The existence of this arbitrariness may
 be exemplified by different extensions of the BV quantization method
 (see, e.g., \cite{multi}).

 The purpose of this report is to demonstrate the fact that extending the
 method of the BV quantization in the part of gauge-fixing, while retaining
 the principles of BRST symmetry and gauge-independence, can lead to off-shell
 different quantum theories. To do so, we consider the superfield form
 \cite{superf} of the BV quantization rules, which provides another
 realization of the principles of BRST symmetry and gauge-independence.
 By extending \cite{superf} in the part of gauge-fixing, we obtain a
 quantization scheme \cite{modsuperf} that differs from a natural
 generalization of the gauge-fixing procedure inherent in the BV quantization.

\section{Batalin--Vilkovisky Formalism}

 Let us remind the basic ingredients of the BV quantization method for general
 gauge theories \cite{BV}.

 The starting point is a classical gauge-invariant action $S_0(A)$ of fields
 $A^i$. The action may belong to the class of irreducible or reducible
 theories, with a closed or open algebra of gauge transformations. The
 complete configuration space $\phi^A$ is introdiced, spanned by the initial
 fields $A^i$, the ghost fields, and the auxiliary fields. The content of the
 complete configuration space is determined by the properties of the initial
 theory, i.e. by the fact whether the theory is an irreducible or reducible
 one. Note that the explicit structure of the complete configuration space is
 not essential for the following considerations.

 With each field $\phi^A$, a corresponding antifield $\phi^*_A$ is accociated,
 having the opposite Grassmann parity,
 $\varepsilon(\phi^*_A)=\varepsilon_A+1$. In the space of the
 field-antifield variables one defines the antisimplectic operation $(\;,\;)$
 called the antibracket
\begin{eqnarray}
 \label{bracket}
 (F,\;G)=\frac{\delta F}{\delta\phi^A}\frac{\delta G}{\delta
 \phi^*_A}-(-1)^{(\varepsilon(F)+1)(\varepsilon(G)+1)}
 \frac{\delta G}{\delta\phi^A}\frac{\delta F}
 {\delta\phi^*_A}
\end{eqnarray}
 and the nilpotent operator $\Delta$
\begin{eqnarray}
 \label{delta}
 \Delta=(-1)^{\varepsilon_A}\frac{\delta_l}{\delta\phi^A}\frac
 {\delta}{\delta\phi^{*}_A}\,,\;\;\;\Delta^2=0.
\end{eqnarray}
 The basic object of the BV quantization is the quantum action
 $S=S(\phi,\phi^*)$ that satisfies the generating equation
\begin{equation}
 \label{BVgeneq1}
 \frac{1}{2}(S,S)=i\hbar\Delta S,
\end{equation}
 or equivalently
\begin{equation}
 \label{BVgeneq2}
 \Delta\exp\bigg\{\frac{i}{\hbar}S\bigg\}=0,
\end{equation}
 with the boundary condition
\[
 S|_{\phi^*=\hbar=0}=S_0.
\]

 The vacuum functional $Z$ is defined as an integral over the fields of the
 complete configuration space $\phi^A$,
\begin{eqnarray}
\label{BVvacfunc}
 Z=\int
 d\phi\;\exp\bigg\{\frac{i}{\hbar}S_{\rm eff}(\phi)\bigg\},
\end{eqnarray}
 where
\begin{eqnarray}
\label{BVeff}
 S_{\rm eff}(\phi)=S_{\rm ext}(\phi,\phi^*)|_{\phi^*=0}\,.
\end{eqnarray}
 In eq.~(\ref{BVeff}), $S_{\rm ext}=S_{\rm ext}(\phi,\phi^*)$ is the
 gauge-fixed quantum action constructed as the symmetry transformation
\begin{eqnarray}
\label{BVoper}
 \exp\bigg\{\frac{i}{\hbar}S_{\rm ext}\bigg\}=
 \exp\bigg(-[\Delta,\Psi]_{+}\bigg)
 \exp\bigg\{\frac{i}{\hbar}S\bigg\},
\end{eqnarray}
 retaining the form of the generating equation (\ref{BVgeneq2})
\begin{equation}
 \label{retain}
 \Delta\exp\bigg\{\frac{i}{\hbar}S_{\rm ext}\bigg\}=0,
\end{equation}
 and introduced with the help of the gauge fermion $\Psi$ that removes
 the degeneracy of the functional integral (\ref{BVvacfunc}).

 Equivalently, the vacuum functional (\ref{BVvacfunc}) can be represented
 in the form of an integral over the variables $\phi^A$, $\phi^*_A$,
 $\lambda_A$,
\begin{eqnarray}
\label{Equiva}
 Z&=&\int d\phi\;d\phi^*\;d\lambda \exp\bigg\{\frac{i}{\hbar}\bigg[
 S_{\rm ext}(\phi,\phi^{*})+\phi^*_A\lambda^A\bigg]\bigg\},
\end{eqnarray}
 where $\lambda_A$ are additional fields with the Grassmann parity opposite
 to that of the fields $\phi^A$, $\varepsilon(\lambda_A)=\varepsilon_A+1$.

 By virtue of eq.~(\ref{retain}), the integrand in eq.~(\ref{Equiva}) is
 invariant under the global supersymmetry transformations
\begin{eqnarray}
 \label{BRSTgen}
 \delta\phi^A=\lambda^A\mu,\;\;\;
 \delta\phi^*_A=\mu\frac{\delta S_{\rm ext}}{\delta\phi^A}\,,\;\;\;
 \delta\lambda^A=0
\end{eqnarray}
 with an anticommuting parameter $\mu$. Eqs.~(\ref{BRSTgen}) realize the
 transformations of BRST symmetry in the space of the variables
 $\phi^A$, $\phi^*_A$, $\lambda_A$.

 It follows from eqs.~(\ref{BVgeneq2}), (\ref{BVoper}) and (\ref{Equiva})
 that the vacuum functional $Z\equiv Z_{\Psi}$ does not depend on the choice
 of the gauge fermion, $Z_{\Psi+\delta\Psi}=Z_\Psi$, and hence, by the
 equivalence theorem \cite{equiv}, the corresponding $S$-matrix is
 gauge-independent.

 When formulating the rules of the BV quantization, the choice of $\Psi$
 in the symmetry transformation (\ref{BVoper}) was made in the form of a
 functional depending on the fields, $\Psi=\Psi(\phi)$. In this case the
 corresponding vacuum functional (\ref{BVvacfunc}) reduces to the well-known
 expression
\begin{eqnarray}
 \label{BV}
 Z=\int d\phi\;d\phi^*\;d\lambda\exp\bigg\{\frac{i}{\hbar}
 \bigg[S(\phi,\phi^*)+\bigg(\phi^*_A-\frac{\delta\Psi}{\delta\phi^A}\bigg)
 \lambda^A\bigg]\bigg\}.
\end{eqnarray}

\section{Modified Superfiled BRST Quantization}

 Recently, a closed superfield form of the BV quantization rules was
 proposed \cite{superf}. In this formalism all variables of the BV approach
 are combined into so-called superfields and super-antifields in a
 superspace spanned by space-time coordinates and a scalar Grassmann
 coordinate. The vacuum functional (\ref{BV}) of the BV formalism is
 contained in the superfield approach as a particular case of gauge-fixing
 and solutions of the generating equation.

 We now generalize \cite{modsuperf} the superfield BRST formalsim
 \cite{superf} in such a way that the gauge is introduced by means of a
 bosonic functional depending on all variables of the superfield formalism,
 including the super-antifields. The gauge-fixing functional is subject to
 a generating equation similar to the equation that determines the quantum
 action.

 Let us introduce a superspace $(x^\mu,\theta)$ spanned by space-time
 coordinates $x^\mu$, $\mu=(0,1,\ldots,D-1)$, and a scalar anticommuting
 coordinate $\theta$. Let $\Phi^A(\theta)$ be a set of superfields
 $\Phi^A(\theta)$, accompanied by a set of the corresponding
 super-antifields $\Phi^*_A(\theta)$ with the Grassmann parities
 $\varepsilon(\Phi^A)\equiv\varepsilon_A,\,
 \varepsilon(\Phi^*_A)=\varepsilon_A+1$. The superfields are subject
 to the boundary condition
\begin{eqnarray}
\label{BondSQ}
 \Phi^A(\theta)|_{\theta =0}=\phi^A.
\end{eqnarray}
 Define the vacuum functional $Z$ as the following integral:
\begin{eqnarray}
 \label{ZSQ}
 Z=\int
 d\Phi\;d\Phi^*\rho[\Phi^*]
\exp\bigg\{\frac{i}{\hbar}\bigg(S[\Phi,\Phi^*] +
X[\Phi,\Phi^*] +\Phi^*\Phi\bigg)\bigg\}.
\end{eqnarray}
 In eq.~(\ref{ZSQ}), $S=S[\Phi,\Phi^*]$ is the quantum action determined by
 the generating equation
\begin{eqnarray}
\label{GEq1SQ}
\frac{1}{2}(S,S)+VS=i\hbar\Delta S,
\end{eqnarray}
 while $X=X[\Phi,\Phi^*]$ is the gauge-fixing functional that satisfies the
 equation
\begin{eqnarray}
\label{GEqXSQ}
\frac{1}{2}(X,X)-UX=i\hbar\Delta X.
\end{eqnarray}

 In eqs.~(\ref{ZSQ}), (\ref{GEq1SQ}) and (\ref{GEqXSQ}), we have used the
 antibracket $(\;,\;)$ expressed in terms of arbitrary functionals
 $F=F[\Phi,\Phi^*]$, $G=G[\Phi,\Phi^*]$ by the rule
\begin{eqnarray}
\label{ABSQ}
 (F,G)&=&\int d\theta\bigg\{\frac{\delta F}{\delta\Phi^A(\theta)}
 \frac{\partial}{\partial\theta}\frac{\delta
 G}{\delta\Phi^*_A(\theta)}(-1)^{\varepsilon_A+1}-\\
\nonumber
 &&-(-1)^{(\varepsilon(F)+1)(\varepsilon(G)+1)}(F\leftrightarrow
 G)\bigg\}.
\end{eqnarray}
 We have also used the operators $\Delta$, $V$ and $U$
\begin{eqnarray}
\label{DeltaSQ}
\Delta&=&-\int d\theta(-1)^{\varepsilon_A}\frac{\delta_l}
 {\delta\Phi^A(\theta)}
 \frac{\partial}{\partial\theta}\frac{\delta}
 {\delta\Phi^*_A(\theta)},\\
\label{VSQ}
 V&=&-\int d\theta\frac{\partial\Phi^*_A(\theta)}{\partial\theta}
 \frac{\delta}{\delta\Phi^*_A(\theta)},\\
\label{USQ}
 U&=&-\int d\theta \frac{\partial\Phi^A(\theta)}
 {\partial\theta}\frac{\delta_l}{\delta\Phi^A(\theta)}
\end{eqnarray}
 (derivatives with respect to $\theta$ are understood as acting from
 the left) as well as the functionals $\rho[\Phi^*]$ and $\Phi^*\Phi$
\begin{eqnarray}
\label{WFuncSQ}
 \rho[\Phi^*]&=&\delta\bigg(\int d\theta\,\Phi^*(\theta)\bigg),\\
 \label{phi_phi*}
 \Phi^*\Phi&=&\int d\theta\,\Phi^*_A(\theta)\Phi^A(\theta).
\end{eqnarray}
 The algebra of the above operators (\ref{DeltaSQ}), (\ref{VSQ}) and
 (\ref{USQ}) reads as follows:
\begin{equation}
 \label{NilSQ}
 \begin{array}{cc}
 \Delta^2=0\;,\;\;U^2=0\;,\;\;V^2=0,&\\
 &\\
 \Delta U+U\Delta=0,\;\;\;\Delta V+V\Delta=0,\;\;UV+VU=0.&
 \end{array}
\end{equation}
 The equations (\ref{GEq1SQ}) and (\ref{GEqXSQ}) can be represented in the
 equivalent form
\begin{equation}
 \label{GEqX1SQ}
\begin{array}{rcl}
 \bar{\Delta}\exp\bigg\{\frac{\textstyle i}{\textstyle\hbar}S\bigg\}&=&0,\\
 \tilde{\Delta}\exp\bigg\{\frac{\textstyle i}{\textstyle\hbar}X\bigg\}&=&0
\end{array}
\end{equation}
 with the help of the operators
\begin{eqnarray*}
 \bar{\Delta}=\Delta + \frac{i}{\hbar}V,\;\;
 \tilde{\Delta}=\Delta - \frac{i}{\hbar}U,
\end{eqnarray*}
 having the algebraic properties
\begin{eqnarray*}
 {\bar{\Delta}}^2=0,\;\; {\tilde{\Delta}}^2=0,\;\;
\bar{\Delta}\tilde{\Delta} + \tilde{\Delta}\bar{\Delta}=0.
\end{eqnarray*}
 By the nilpotency (\ref{NilSQ}) of the operator $U$ (\ref{USQ}), any
 functional
\begin{equation}
 \label{XU}
 X=U\Psi[\Phi]
\end{equation}
 with a fermionic functional $\Psi[\Phi]$ is a solution of the equation
 (\ref{GEqXSQ}). The above expression (\ref{XU}) has the exact form of the
 gauge-fixing functional applied by the method of superfield BRST
 quantization \cite{superf}.

 By virtue of eqs.~(\ref{GEq1SQ}) and (\ref{GEqXSQ}), the integrand in
 eq.~(\ref{ZSQ}) is invariant under global supersymmetry transformations
 with an anticommuting parameter $\mu$,
\begin{equation}
\begin{array}{lcr}
 \label{BRSTSQ}
 \delta\Phi^A(\theta)&=&\mu U\Phi^A(\theta)+(\Phi^A(\theta),X-S)\mu,
 \\
 \delta\Phi^*_A(\theta)&=&\mu V\Phi^*_A(\theta)+(\Phi^*_A(\theta),X-S)\mu.
\end{array}
\end{equation}
 Eqs.~(\ref{BRSTSQ}) are the transformations of BRST symmetry in the
 framework of the modified superfield BRST quantization.

 By virtue of eqs.~(\ref{GEq1SQ}), (\ref{GEqXSQ}), (\ref{NilSQ}) and
 (\ref{BRSTSQ}), the vacuum functional $Z\equiv Z_X$ in eq.~(\ref{ZSQ})
 does not depend on the choice of the gauge boson, $Z_{X+\delta X}=Z_X$,
 which, consequently, ensures \cite{equiv} the gauge-independence of the
 $S$-matrix.

 The method of modified superfield BRST quantization permits to generalize
 the BV quantization scheme with respect to the gauge-fixing procedure. In
 fact, consider the component representation of the superfields
 $\Phi^A(\theta)$ and super-antifields $\Phi^*_A(\theta)$
\begin{eqnarray}
\nonumber
 \Phi^A(\theta)= \phi^A + \lambda^A\theta,\quad
 \Phi^*_A(\theta)=\phi^*_A - \theta J_A.
\end{eqnarray}
 The set of the variables $\phi^A$, $\phi^*_A$, $\lambda^A$ and $J_A$
 coincides with the complete set of variables applied by the BV method, where
 the components $J_A$, $\varepsilon(J_A)=\varepsilon_A$, are identified with
 the sources to the fields $\phi^A$.

 In the component form, the antibracket (\ref{ABSQ}) and the operator
 $\Delta$ (\ref{DeltaSQ}) coincide with the corresponding objects
 (\ref{bracket}) and (\ref{delta}) of the BV method.

 Let us now represent the integration measure (\ref{WFuncSQ}) in the
 component form
\begin{eqnarray*}
 d\Phi\;d\Phi^*\;\rho[\Phi^*]=d\phi\;d\phi^*\;d\lambda\;dJ\;\delta(J).
\end{eqnarray*}
 Solutions of the generation equation that determine the action $S$ when
 $J_A=0$ may be sought among solutions of the master equation (\ref{BVgeneq1})
 applied by the BV method, since the operator $V$ (\ref{VSQ})
\begin{eqnarray*}
 V=-J_A\frac{\delta}{\delta\phi^*_A}
\end{eqnarray*}
 vanishes when $J_A=0$.

 Let us choose the functional $S$ to be independent of $\lambda^A$. Taking
 into account the component form of $\Phi^*\Phi$ in eq.~(\ref{phi_phi*})
\begin{eqnarray*}
\Phi^*\Phi=\phi^*_A\lambda^A-J_A\phi^A,
\end{eqnarray*}
 we obtain the following representation of the vacuum functional (\ref{ZSQ}):
\begin{eqnarray}
\label{GFmodBV}
 Z=\int d\phi\;d\phi^*\;d\lambda\exp\bigg\{\frac{i}{\hbar}
 \bigg[S(\phi,\phi^*)+X(\phi,\phi^*,\lambda)+\phi^*_A\lambda^A\bigg]\bigg\}.
\end{eqnarray}

 The above result (\ref{GFmodBV}) may be considered as a natural extension
 of the BV quantization procedure to a more general case of gauge-fixing.
 In fact, the functional $X=U\Psi[\Phi]$ is a solution of the generating
 equation (\ref{GEqXSQ}). Note that the component representation of the
 operator $U$ (\ref{USQ}) reads
\begin{eqnarray*}
 U=-(-1)^{\varepsilon_A}\lambda^A\frac{\delta_l}{\delta \phi^A}\;.
\end{eqnarray*}
 Let us choose the functional $\Psi$ to be independent of the fields
 $\lambda^A$, $\Psi=\Psi(\phi)$. Then we find that the gauge-fixing
 functional $X$
\begin{eqnarray}
 \label{Xphi}
 X(\phi,\lambda)=-\frac{\delta\Psi(\phi)}{\delta \phi^A}\lambda^A
\end{eqnarray}
 becomes identical with the gauge applied by the BV quantization method,
 and therefore in the above particular case (\ref{Xphi}) of solutions that
 determine $X$ the vacuum functional (\ref{GFmodBV}) of the modified
 superfiels BRST formalism coincides with that of the BV method (\ref{BV}).
 This means, given the established gauge-independence within the BV approach
 as well as within the modified superfield BRST formalism, the coincidence
 of the $S$-matrices in both these methods, including the case when the
 gauge-fixing of the BV method is extended to an arbitrary (admissible) gauge
 fermion $\Psi$ in the symmetry transformation (\ref{BVoper}).

 Let us show, however, that the extension of the gauge-fixing procedure of
 the BV method obtained as a generalization of the gauge fermion does not
 coincide with that provided by the modified superfield BRST formalism.
 Consider the generalization of the BV vacuum functional (\ref{BV}) following
 from eq.~(\ref{BVvacfunc}) with the symmetry transformation (\ref{BVoper})
 extended to the case of the gauge fermion chosen as a functional of fileds
 and antifields, $\Psi=\Psi(\phi,\phi^*)$. Then the original vacuum functional
 (\ref{BVvacfunc}) can be represented in the form
\begin{eqnarray}
 \label{ZBV2}
 Z=\int d\phi\;d\phi^*\;d\lambda\exp\bigg\{\frac{i}{\hbar}
 \bigg[S(\phi,\phi^*)+S_{\Psi}(\phi,\lambda)+\phi^*_A\lambda^A\bigg]\bigg\},
\end{eqnarray}
 where the corresponding gauge-fixed part $S_{\Psi}(\phi,\lambda)$ of the
 quantum action,
\begin{eqnarray*}
 \exp\bigg\{\frac{i}{\hbar}S_{\Psi}(\phi,\lambda)\bigg\}&=&\int
 d\phi{'}\;d\phi^{*'}\;\delta\bigg(\phi-\phi{'}
 +\frac{\delta\Psi}{\delta\phi^{*'}}\bigg)\delta(\phi^{*'})\times\\
 &&\times\exp\bigg\{\frac{i}{\hbar}\bigg[-\bigg(\phi^{*'}
 +\frac{\delta\Psi}{\delta\phi{'}}\bigg)\lambda
 +i\hbar\Delta{'}\Psi\bigg]\bigg\},
\end{eqnarray*}
 does not depend on the antifields. In this sense, the introduction of
 gauge with the help of the bosonic functional $X$ (\ref{GEqXSQ}) and
 (\ref{GFmodBV}) in the framework of the modified superfield BRST approach
 can be considered as a formal extension of the gauge-fixing procedure
 introduced by means of the symmetry transformation (\ref{BVoper}) within
 the BV quanization method.

 Note that gauge-fixing in terms of the bosonic functional $X$ (\ref{GEqXSQ}),
 or (\ref{GEqX1SQ}), aslo leads to a generalization of the usual Ward
 identities. Indeed, define, as we return to the vacuum functional
 (\ref{GFmodBV}), the extended generating functional $Z(J,\phi^*)$ of Green's
 functions by the rule
\begin{eqnarray}
 \label{GFmodGreen}
 Z(J,\phi^*)&=&\int
 d\phi{'}\;d\phi^{*'}\;d\lambda\exp\bigg\{\frac{i}{\hbar}
 \bigg[S(\phi{'},\phi^{*'})+X(\phi{'},\phi^{*'},\lambda)\nonumber\\
 &&+\,(\phi^{*'}-\phi^*)\lambda+J\phi{'}\bigg]\bigg\},
\end{eqnarray}
 where we have introduced the sources $J_A$ to the fields $\phi^A$.

 In the particular case when the bosonic functional $X$, by analogy with
 the BV method (\ref{ZBV2}), does not contain dependence on the antifields,
 $X=X(\phi,\lambda)$, the generating equation (\ref{GEqX1SQ}) that
 determines $X$ takes on the form
\begin{eqnarray}
\label{GEqXsimp}
 U\exp\bigg\{\frac{i}{\hbar}X\bigg\}=0.
\end{eqnarray}
 From eqs.~(\ref{GFmodGreen}) and (\ref{GEqXsimp}), with allowance for
 eq.~(\ref{BVgeneq1}), follow the well-known Ward identities for gauge
 theories
\begin{eqnarray*}
 J_A\frac{\delta}{\delta\phi^*_A}Z(J,\phi^*)=0.
\end{eqnarray*}
 When extending eq.~(\ref{GEqXsimp}) to the case (\ref{GEqX1SQ}) that
 corresponds to the presence of antifields in the gauge-fixing functional,
 $X=X(\phi,\phi^*,\lambda)$, the form of the corresponding Ward identities
 is modified to
\begin{eqnarray*}
 J_A\frac{\delta}{\delta\phi^*_A}Z=\frac{1}{i\hbar}J_A
 \left<\frac{\delta X}{\delta\phi^{*'}_A}\right>Z,
\end{eqnarray*}
 where we have used the notation $\left<\delta X/\delta\phi^{*'}_A\right>$
 for the vacuum expectation value of the functional
 $\delta X/\delta\phi^{*'}_A$\,, $X=X(\phi{'},\phi^*{'},\lambda)$.

\section{Conclusion}
 In this report, we have considered the modified scheme \cite{modsuperf} of
 superfield Lagrangian quantization based on the principles of BRST symmetry
 and the gauge-independence of the $S$-matrix. We have shown that the
 formalism \cite{modsuperf} provides a generalization of the gauge-fixing
 procedure of the BV quantization method \cite{BV}, different from the
 natural extension of the gauge-fixing inherent in \cite{BV}. Thus we
 conclude that the above two extensions of the BV approach provide off-shell
 different realizations of the principles of BRST symmetry and
 gauge-independence, while leading to identical $S$-matrices.\\

\noindent{\bf Acknowledgements}
 The work was partially supported by a grant of the Ministry of General and
 Professional Education of the Russian Federation in the field of fundamental
 sciences, as well as by grant RFBR 99-02-16617. The work of P.M.L. was also
 partially supported by grants INTAS 991-590 and RFBR-DFG 99-02-04022.

\end{document}